\documentstyle{mn}

\newif\ifAMStwofonts

\ifoldfss
  \ifCUPmtlplainloaded \else
    \NewTextAlphabet{textbfit} {cmbxti10} {}
    \NewTextAlphabet{textbfss} {cmssbx10} {}
    \NewMathAlphabet{mathbfit} {cmbxti10} {} 
    \NewMathAlphabet{mathbfss} {cmssbx10} {} 
  \fi
  \ifAMStwofonts
    \ifCUPmtlplainloaded \else
      \NewSymbolFont{upmath} {eurm10}
      \NewSymbolFont{AMSa} {msam10}
      \NewMathSymbol{\upi}     {0}{upmath}{19}
      \NewMathSymbol{\umu}     {0}{upmath}{16}
      \NewMathSymbol{\upartial}{0}{upmath}{40}
      \NewMathSymbol{\leqslant}{3}{AMSa}{36}
      \NewMathSymbol{\geqslant}{3}{AMSa}{3E}

       \let\le=\leqslant
       
    \fi
  \fi
\fi 

\ifnfssone
  \newmathalphabet{\mathit}
  \addtoversion{normal}{\mathit}{cmr}{m}{it}
  \addtoversion{bold}{\mathit}{cmr}{bx}{it}
  \newmathalphabet{\mathbfit} 
  \addtoversion{normal}{\mathbfit}{cmr}{bx}{it}
  \addtoversion{bold}{\mathbfit}{cmr}{bx}{it}
  \newmathalphabet{\mathbfss} 
  \addtoversion{normal}{\mathbfss}{cmss}{bx}{n}
  \addtoversion{bold}{\mathbfss}{cmss}{bx}{n}
  \ifAMStwofonts
    \ifCUPmtlplainloaded \else
      \UseAMStwoboldmath
      \makeatletter
      \new@mathgroup\upmath@group
      \define@mathgroup\mv@normal\upmath@group{eur}{m}{n}
      \define@mathgroup\mv@bold\upmath@group{eur}{b}{n}
      \edef\UPM{\hexnumber\upmath@group}
      \new@mathgroup\amsa@group
      \define@mathgroup\mv@normal\amsa@group{msa}{m}{n}
      \define@mathgroup\mv@bold\amsa@group{msa}{m}{n}
      \edef\AMSa{\hexnumber\amsa@group}
      \makeatother
      \mathchardef\upi="0\UPM19
      \mathchardef\umu="0\UPM16
      \mathchardef\upartial="0\UPM40
      \mathchardef\leqslant="3\AMSa36
      \mathchardef\geqslant="3\AMSa3E

       \let\le=\leqslant

    \fi
  \fi
\fi 

\ifnfsstwo
  \DeclareMathAlphabet{\mathbfit}{OT1}{cmr}{bx}{it}
  \SetMathAlphabet\mathbfit{bold}{OT1}{cmr}{bx}{it}
  \DeclareMathAlphabet{\mathbfss}{OT1}{cmss}{bx}{n}
  \SetMathAlphabet\mathbfss{bold}{OT1}{cmss}{bx}{n}
  \ifAMStwofonts
    \ifCUPmtlplainloaded \else
      \DeclareSymbolFont{UPM}{U}{eur}{m}{n}
      \SetSymbolFont{UPM}{bold}{U}{eur}{b}{n}
      \DeclareSymbolFont{AMSa}{U}{msa}{m}{n}
      \DeclareMathSymbol{\upi}{0}{UPM}{"19}
      \DeclareMathSymbol{\umu}{0}{UPM}{"16}
      \DeclareMathSymbol{\upartial}{0}{UPM}{"40}
      \DeclareMathSymbol{\leqslant}{3}{AMSa}{"36}
      \DeclareMathSymbol{\geqslant}{3}{AMSa}{"3E}

       \let\le=\leqslant

    \fi
  \fi
\fi 

\ifCUPmtlplainloaded \else
  \ifAMStwofonts \else 
    \def\upi{\pi}
    \def\umu{\mu}
    \def\upartial{\partial}
  \fi
\fi

\title{Gamma-rays and cosmic-rays from a pulsar in Cygnus OB2} 
\author[W. Bednarek]
       {W. Bednarek \\
        Department of Experimental Physics, University of \L \'od\'z,
        ul. Pomorska 149/153, 90-236 \L \'od\'z, Poland}
\date{Accepted 
      Received ;
      in original form }
\pubyear{}

\begin{document}

\maketitle

\begin{abstract}
We argue that gamma-ray sources observed in the direction of the Cyg OB2 
association in the GeV and TeV energy range are due to a pulsar which was 
created by a supernova a few tens of thousands years ago. The GeV emission
is produced by a middle-aged pulsar, a factor of two older than the Vela pulsar.
The TeV emission is produced by high energy hadrons and/or leptons accelerated 
in the pulsar wind nebula.
We suggest, moreover, that the excess of cosmic rays at $\sim 10^{18}$ eV, 
observed from the direction of the Cygnus region, can also be related to the appearance 
of this very energetic pulsar in the Cyg OB2 association. A part of relativistic hadrons, 
captured in strong magnetic fields of a high density region of the Cyg OB2,
produce neutrons and $\gamma$-rays in collisions with the matter.   
These neutrons can arrive from the Cyg OB2 creating an excess of cosmic rays.

\end{abstract}


\section{Introduction}

The Cygnus OB2 is the largest massive galactic association at a relatively small distance 
of 1.7 kpc. It has the diameter of $\sim 60$ pc and the core radius of $\sim 14\pm 2$ pc.
The total H$_2$ mass of the Cyg OB2 derived from the CO survey is $\sim 3.3\times 10^5$ 
M$_{\odot}$ (Butt et al.~2003), and the total stellar mass is $(4 - 10)\times 
10^4$ M$_{\odot}$ (Kn\"odlseder~2000). It contains many massive stars of the OB type 
($\sim 2600\pm 400$) and O type ($\sim 120\pm 20$) (Kn\"odlseder~2000). The typical 
main sequence evolution lifetime of the massive O type stars is of the order of a few 
$10^6$ yrs and B type stars a few tens of $10^6$ yrs. Massive O type stars should 
pass through the Wolf-Rayet phase and explode 
as supernovae. It is expected that very fast pulsars are created in explosions of 
such massive stars at a rate of about one per a few $10^4$ yrs within the Cyg OB2. This 
rate can be much higher if most of these massive stars are of similar age. 
The pulsars, created $\sim 10^4$ yrs ago, should resemble the Vela type 
pulsars (e.g. Vela pulsar, PSR 1706-44), which are strong sources of 
pulsed $\gamma$-rays in the hundred MeV-GeV energies (EGRET observations, Kanbach et 
al.~1994, Thompson et al.~1996). These pulsars are also surrounded by relatively 
small pulsar wind nebulae (PWNe), which are sources of TeV $\gamma$-rays 
(Chadwick et al.~2000, Kifune et al.~1995, Chadwick et al.~1998).

The Cygnus region is difficult to observe in high energies due to its very complex 
structure and location on the galactic plane. The analysis of the EGRET data from 
different viewing periods revealed a few sources in a relatively
small field of view containing the Cyg OB2:  2EG J2033+4112 (Mori et al.~1997), 
GeV J2035+4214 (Lamb \& Macomb~1997), GRO J2034+4203 (Reimer, Dingus \& Nolan~1997),
3EG J2033+4118 (Hartman et al.~1999). The first source is interpreted by the authors as 
caused by the massive binary Cyg X-3. The other sources may be due to the same object
in the Cyg OB2, according to Butt et al.~(2003). 

Recently, the HEGRA group has reported the discovery of an unidentified, steady, 
TeV source at the 
edge of the 95\% error circle of the source 3EG J2033+4118 (Aharonian et al.~2002). 
The spectrum of this source is rather flat (differential spectral index $-1.9\pm 
0.3_{\rm stat}\pm 0.3_{\rm sys}$), with the flux about two orders below the flux
observed from 3EG J2033+4118. Moreover, there is an indication that the TeV source is
extended with a radius of $5.6'$ (Aharonian et al.~2002), confirmed in the new analysis 
of the HEGRA data by Rowell \& Horns~(2002). 

The discovery of TeV source has led to a more careful analysis of the X-ray 
observations of Cyg OB 2 from the ROSAT and Chandra telescopes (Mukherjee et al.~2003, 
Butt et al.~2003). None of the several X-ray sources detected by the Chandra in the region 
with $\sim 11'$ diameter around the TeV source was particularly prominent (Butt et 
al.~2003). Mukherjee et al.~(2003) notes only one interesting variable source which 
lies $7'$ from the TeV centroid with the flux of $2\times 10^{-12}$ ergs cm$^{-2}$ s$^{-1}$.
It might be a quiescent low-mass X-ray binary 
containing a neutron star or a black hole. However, weak diffuse X-ray emission has 
also been observed, with the intensity of $1.3\times 10^{-12}$ ergs cm$^{-2}$ s$^{-1}$ in 
0.5-2.5 keV and $3.6\times 10^{-12}$ ergs cm$^{-2}$ s$^{-1}$ in 2.5-10 keV 
(Butt et al.~2003). This is more than ten times as bright as the sum of all the 
point-like sources. 
 
Recent analysis of the arrival directions of cosmic rays with energies of $\sim 10^{18}$ eV
by the AGASA group shows the excess of particles from the direction of the Galactic 
Center (Hayashida et al.~1999a, confirmed by the SUGAR analysis, Bellido et al.~2001),
and from the direction of the Cygnus region. The second excess has a significance of 
3.9 $\sigma$ (3401 events with the background of 3148) in the energy range 
$10^{18} - 10^{18.4}$ eV (Hayashida et al.~1999b). We suggest that this excess of EHE 
particles can also be related to the high energy processes occurring in the massive 
Cyg OB2 association.

The possibility that $\gamma$-rays are produced by massive young stellar clusters, 
as a result of acceleration of particles in the winds or shocks produced by the early 
type stars, has been frequently discussed in the past (with the application to the 
Cyg OB2, see e.g. Chen, White \& Bertsch~1992, Manchanda et al.~1996, 
Benaglia et al.~2001). Aharonian et al.~(2002) discusses such general scenario,
suggesting that the TeV emission may originate in collisions of hadrons with a dense 
cloud. These authors also propose that $\gamma$-ray emission may originate in the 
inverse Compton scattering by 
electrons, which are accelerated by a jet-driven termination shock expelled from a 
microquasar, e.g. Cyg X-3 (see Aharonian \& Atoyan~1998). Butt et al.~(2003) argue
that the TeV emission is more likely to originate in collisions of hadrons with the 
matter than in the leptonic processes. Hadrons might be accelerated to PeV energies in 
collisions of strong stellar winds. 

In this paper we interpret the variety of high energy processes connected with the 
Cyg OB2 association by applying the model in which the basic role is played by a young 
pulsar, born in this association about $\sim 10^4$ yrs ago. It is shown that the 
observed GeV emission can be due to the Vela type pulsar present inside or close to the
Cyg OB2 (Sect.~3). The TeV $\gamma$-ray emission originates in the pulsar wind nebula as 
a result of interactions of hadrons and leptons (Sect.~4). The observed excess of cosmic 
rays from the direction of the Cygnus region at energies of $\sim 10^{18}$
eV is due to neutrons, which are dissolved from heavy nuclei in their interactions 
with the matter of the Cyg OB2 (Sect.~5). We show that if a small part of nuclei, which 
escaped from the PWNa, is captured in a strong magnetic field of dense regions of the 
Cyg OB2 association, the flux of neutrons can explain the observed cosmic ray 
excess from the Cygnus region. 

\section{A model for high energy processes in the Cyg OB2}

Let us assume that a very energetic pulsar is formed in a core collapse of one of 
the massive stars in the Cyg OB2 association. The massive O type stars should evolve
through the Wolf-Raeyt phase during which the star loses  a significant part of its 
envelope, rotates fast, and has strong surface magnetic fields. The Wolf Raeyt stars 
are probably progenitors of neutron stars with extreme parameters (period, surface 
magnetic field) which allow acceleration of particles to very high energies.
Following the recent works (e.g. Gallant \& Arons~1994, Blasi, Epstein \& Olinto~2000) 
we assume that such neutron stars can accelerate
heavy nuclei in their wind zones (e.g. in the mechanism called magnetic slingshot, 
Gunn \& Ostriker~1969). Arons and collaborators (see e.g. Arons~1998) argue
that the iron nuclei can reach  
energies equal to a significant part, $\chi$, of the total potential drop through the 
pulsar polar cap region,
\begin{eqnarray}
E_{\rm Fe} = \chi Z e \Phi_{\rm open},
\end{eqnarray}
\noindent
where, $Z e$ is the charge of the iron nuclei, $\Phi_{\rm open} = 
\sqrt{E_{\rm rot}/c}$ is the maximum potential drop across the polar cup of the pulsar, 
$E_{\rm rot}$ is the rate of rotational energy lost by the pulsar, 
$\chi\le 1$, and $c$ is the velocity of light. These nuclei are accelerated with 
efficiency corresponding to the Goldreich \& Julian current at the light cylinder radius 
(Goldreich \& Julian~1969). Due to the rotational energy losses of the pulsar, the 
energies of nuclei injected into the expanding nebula drop with time.
Based on the modelling of the Crab Nebula, Arons (1998) argues that $\chi$ should be
not very far from one. Assuming $\chi = 0.5$, we use the following prescription for the 
evolution of the Lorentz factor of injected Fe nuclei with time 
\begin{eqnarray}
\gamma_{\rm Fe} = E_{\rm Fe}/m_{\rm Fe}\approx  4\times 10^9 B_{12} P_{\rm ms}^{-2},
\end{eqnarray}
\noindent
where $m_{\rm Fe}$ is the mass of the iron nuclei, $B = 10^{12}B_{12}$ G is the surface 
magnetic field of the pulsar, and $P = 10^{-3}P_{\rm ms}$ s is the pulsar period 
which evolves in time $t$ (in seconds) due to the rotational energy losses according to 
\begin{eqnarray}
P_{\rm ms}^2 = P_{\rm 0, ms}^2 + 2\times 10^{-9} t B_{12}^2,
\end{eqnarray}
\noindent
where $P_{\rm 0, ms}$ is an initial period of the pulsar.

In order to obtain the 
equilibrium spectra of these nuclei inside the nebula at a specific time, we construct
a simple time dependent model for the expanding nebula, basing on the early 
papers which discuss the evolution of a supernova envelope under the influence of a 
young pulsar (e.g. Ostriker \& Gunn~1971). This model takes into account possible
acceleration of the nebula due to the energy supply by the pulsar. This has influence
on the main parameters 
of the nebula, i.e. an expansion velocity, a shock radius, an outer radius, and 
a magnetic field inside the nebula. The details of such a model are described in 
Sect.~2 of Bednarek \& Bartosik~(2003). The equilibrium spectra of nuclei are 
obtained by integrating their injection spectra at a specific time  
over the activity period of the pulsar, and taking into account different processes 
such as: (1) inelastic collisions of nuclei with the matter (important at an early phase); 
(2) the adiabatic energy losses of particles due to expansion of the nebula; and (3) the 
escape of nuclei from the nebula. We include all these processes as it has been already 
done in our previous papers (see sect.~4 in Bednarek \& Protheroe~2002). 
However, in order not to complicate the model too much, the gravitational energy losses
of the pulsar are neglected. We also include partial disintegration of iron nuclei, 
extracted from the neutron star surface, during their propagation through 
the radiation field of the pulsar outer gaps in the inner pulsar magnetosphere following
Bednarek \& Protheroe~(1997). Even for the pulsars with the age $2\times 10^4$ years, 
the iron nuclei still lose on average one nucleon in collisions with the 
nonthermal radiation in the pulsar's outer gap. Nucleons (protons, neutrons) dissolved 
from the iron nuclei in this process are added to the spectrum of nuclei injected
into the nebula. The example equilibrium spectrum of nuclei, which are still inside the 
nebula at $2\times 10^4$ years after supernova explosion, is shown in Fig.~1. A higher 
energy bump in the spectrum of hadrons contains heavy nuclei of the iron group. Their 
energies are limited to a relatively narrow range due to the 
adiabatic energy losses of nuclei injected at the earlier phase and the escape of more
energetic nuclei from the nebula. A lower energy bump in the spectrum of hadrons 
contains protons from disintegrations of heavy nuclei during their propagation in the 
pulsar outer gap and from collisions of nuclei with a dense matter of the nebula 
at an early stage after supernova explosion. All the medium mass nuclei with relatively 
large Lorentz factors, produced in collisions of nuclei with the matter of the nebula at 
an early phase after the explosion, have already managed to escape from the nebula.  
The equilibrium spectra of nuclei inside the nebula at a specific time allow us to
calculate the spectra of $\gamma$-rays from decay of pions,
produced by nuclei in collisions with the matter of the nebula. 

\begin{figure}
  \vspace{6.truecm}
\includegraphics{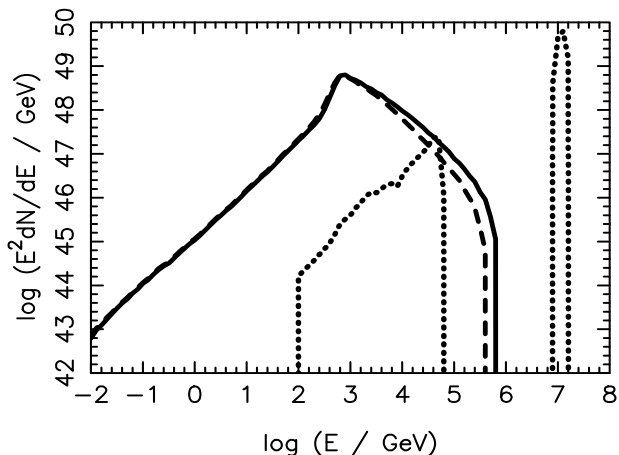}
  \caption{The equilibrium spectra of leptons and different types of nuclei 
(dotted curve) inside the nebula at the time
$2\times 10^4$ yrs after supernova explosion. The initial period of the pulsar
is 2 ms and its surface magnetic field $6\times 10^{12}$ G, and the initial expansion 
velocity of the supernova remnant is 2000 km s$^{-1}$, and its mass is 4 M$_\odot$.
The spectra of leptons are shown for two radiation fields inside the nebula:
(1) the synchrotron photons (SYN) and the microwave background radiation (MBR) 
(full curve), and (2) SYN, MBR, and the infrared photons with temperature 100 K and 
energy density 5 times higher than the MBR (dashed curve).} 
\label{fig1}
\end{figure}

According to Hoshino et al.~(1992), relativistic heavy nuclei can transfer a part of their 
energy, $\xi$, to relativistic positrons in the process of resonant scattering in 
the region above the pulsar wind shock. As a result, positrons are accelerated with
the power law spectrum with the index $\alpha \sim 2$ between 
$m_{\rm e}c^2\gamma_{\rm Fe}$ and $\gamma_{\rm Fe}m_{\rm Fe}c^2/Z$, where  
$m_{\rm e}$ is the mass of the electron. 
The injection spectra of these positrons evolve in time due to the change of energies 
and injection rate of iron nuclei with the pulsar period. The equilibrium spectra
of positrons inside the nebula at a specific time are calculated by taking into account
their radiation energy losses on bremsstrahlung, synchrotron, and inverse Compton (IC) 
processes, and due to the adiabatic losses caused by an expansion of the nebula as it was 
shown in Sect.~4.2 in Bednarek \& Bartosik~(2003). 
The efficiencies of specific energy loss processes are determined by the evolving 
in time parameters of the nebula. The bremsstrahlung 
process depends on the average density of matter inside the nebula which can be easily
evaluated for the known total mass of the nebula and its radius at a specific time.
The synchrotron losses of leptons are determined by the magnetic field 
strength inside the nebula which is estimated from the known dependence of the magnetic 
field inside the pulsar inner magnetosphere and the pulsar wind zone
(see for details Eq.~7 in sect.~2 in Bednarek \& Bartosik~2003).
The evolution of the magnetic field strength in the pulsar wind postshock region   
is shown in Fig.~2b, for the example parameters of the pulsar and the nebula.
The IC losses of leptons is more difficult to estimate. For relatively old nebulae
the energy density of background radiation is dominated by the microwave background 
radiation (MBR),
and possibly infrared background and optical background. Since the optical background 
created by the massive stars inside the Cyg OB2 becomes important only for low energy 
leptons, $\le 100$ GeV, we do not consider their influence on the equilibrium spectrum of 
leptons inside the nebula. The radiation field inside relatively young nebulae is
dominated by the synchrotron photons produced by leptons inside the nebula 
which spectrum is determined by the conditions inside the nebula in the past.
Therefore, in order to estimate the density of synchrotron photons 
at a specific time we apply simple model for the evolution of nebula and include the 
above defined prescription for the injection of particles and the evolution of the 
radiation processes. 
It is assumed that only a part of the whole nebula radiate efficiently. This might be 
motivated by the observed anisotropic injection of particles into the pulsar
wind nebulae which manifests itself by the presence of X-ray torus in the equatorial 
plane of the pulsars (e.g. the X-ray torus around $\sim 20^o$ of the equatorial plane  
observed in the Crab Nebula). 
We estimate the density of synchrotron photons inside the 
nebula assuming that leptons are injected and radiate only within the part of the nebula
limited by the above mentioned $20^o$ torus. The evolution of the synchrotron background 
inside the nebula at a specific time after supernova explosion is shown in 
Fig.~2c. It is clear that inside the nebulae older than
$\sim 3\times 10^3$ yrs, the IC energy losses of leptons on synchrotron photons
becomes negligible. Only scattering of the MBR and possibly the infrared radiation, 
produced by the hot gas inside the nebula, plays important role.
Fig.~2a shows the equilibrium spectra of leptons inside the nebula at a specific time 
after supernova explosion. The lower energy part of the spectrum of leptons 
is formed by leptons which were injected relatively early after supernova explosion
and lose energy at latter time mainly due to the adiabatic expansion of the nebula. 
Therefore, the low energy cut-off in the spectrum of leptons, present
between 0.1 - 0.01 GeV (Fig.~2a), shifts to lower energies
with the evolution of the nebula. The efficient synchrotron cooling determines 
the shape of the spectrum in the higher energy range when the nebula is young and 
the magnetic field is strong. This process is responsible for the 
power low spectrum with the index close to 2 (seen in Fig.~2a) at the time $10^2$ yrs.
On the other hand, the spectrum of leptons at higher energy part at the latter phase 
of expansion of the nebula is 
formed by freshly injected leptons which are not able to cool
to low energies fast enough due to the weak magnetic and photon fields inside the nebula.

\begin{figure}
  \vspace{17.truecm}
\includegraphics{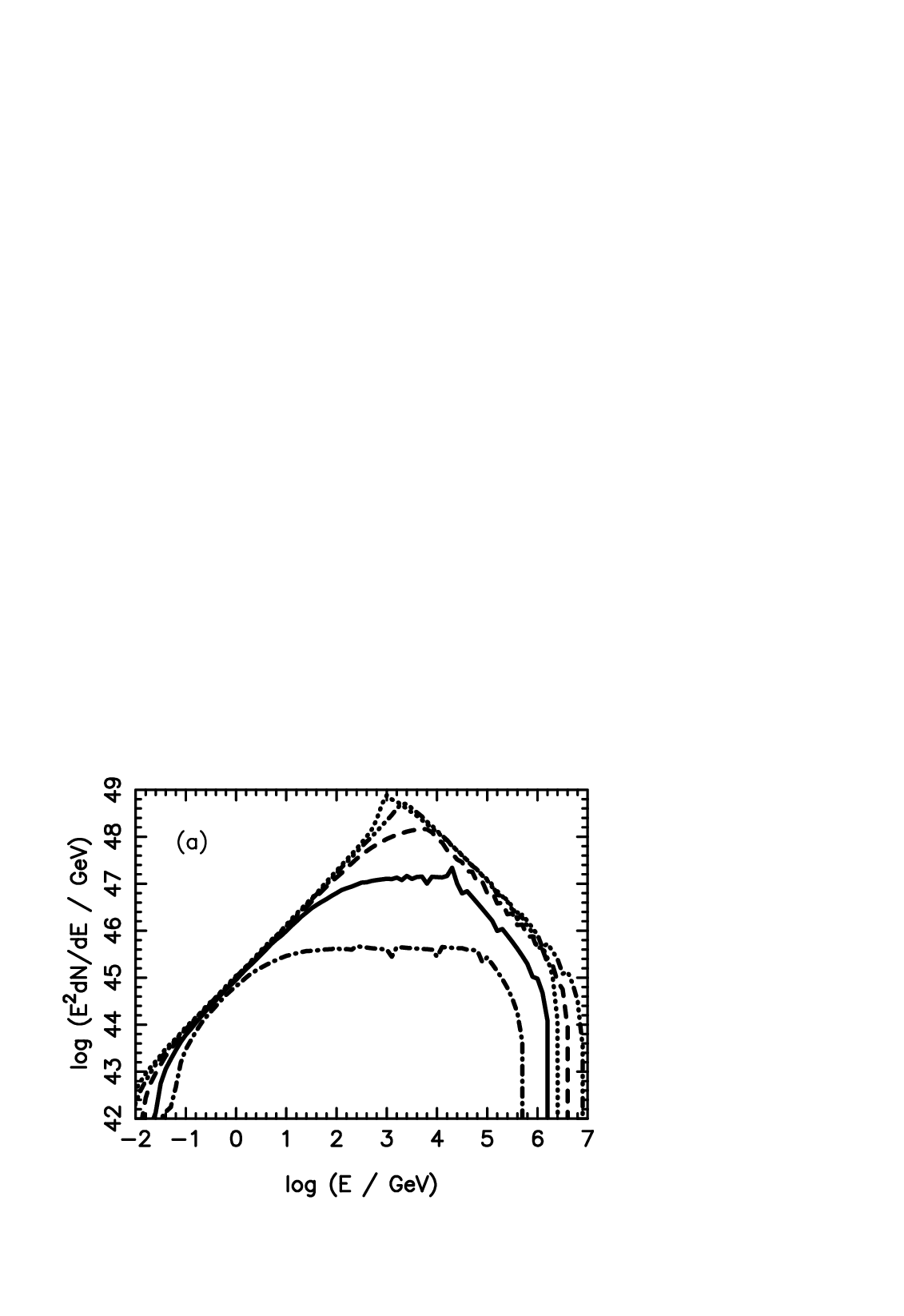}
\includegraphics{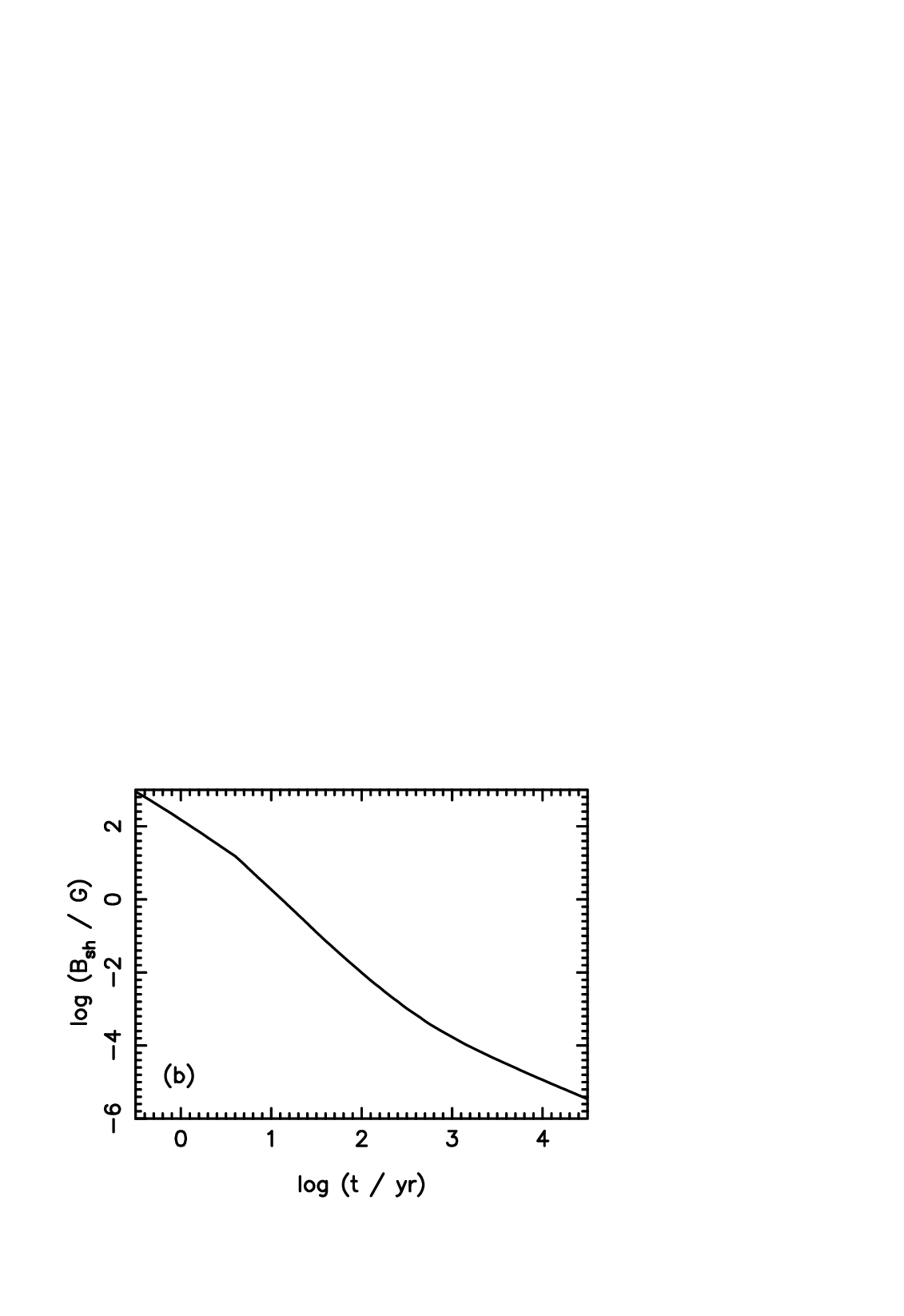}
\includegraphics{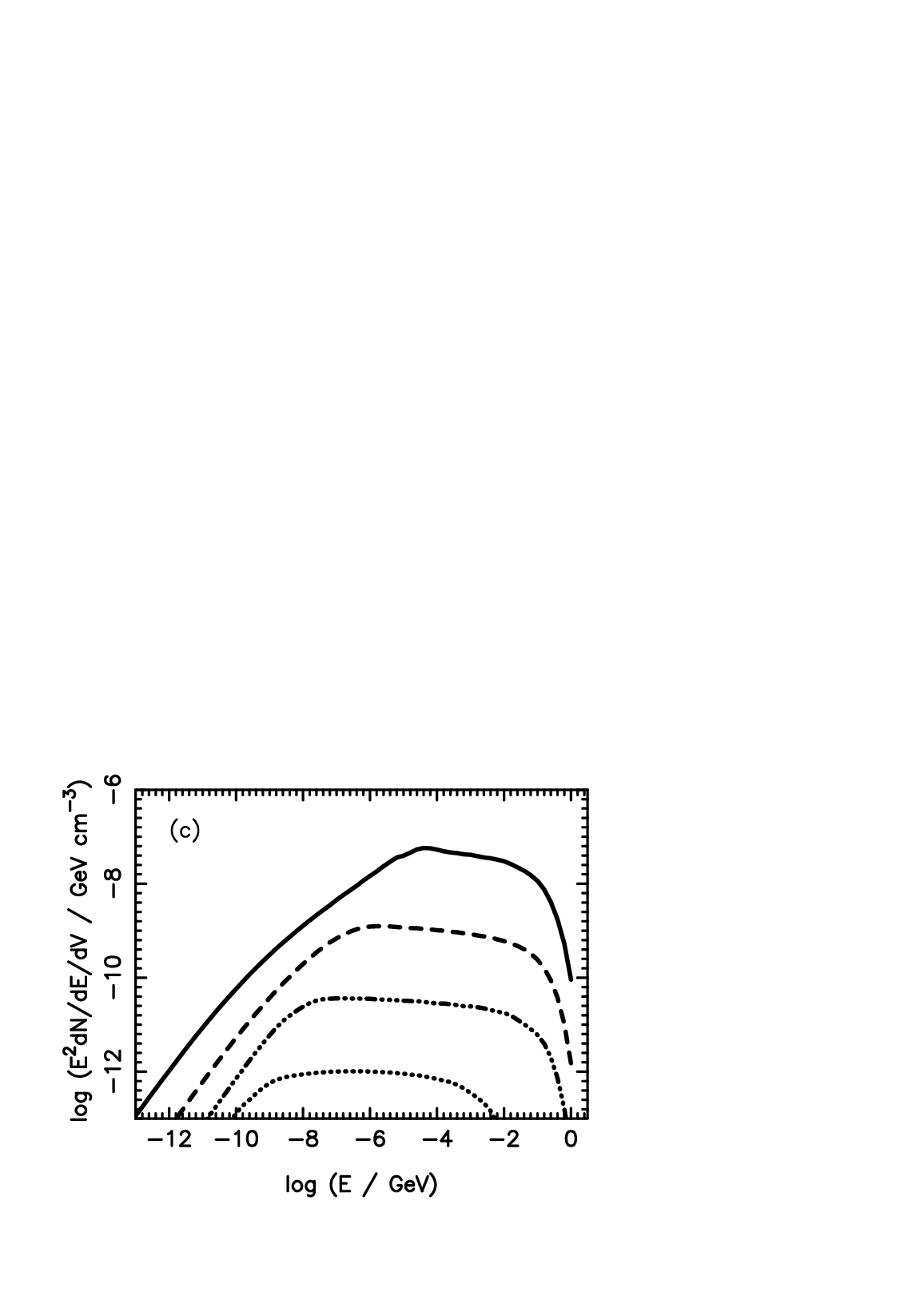}
  \caption{(a) The equilibrium spectra of leptons inside the nebula, injected
with the power law spectrum as discussed in the main text, 
calculated for the pulsar with an initial period of 2 ms and a surface magnetic field of
$6\times 10^{12}$ G, and the supernova with a mass of $4 M_\odot$, expanding with an 
initial velocity of 2000 km s$^{-1}$, are show for different time after a supernova 
explosion (and a pulsar formation) $10^2$ yrs (dot-dashed curve), $3\times 10^2$ yrs 
(full), $10^3$ yrs (dashed), $3\times 10^3$ yrs (dot-dot-dot-dashed), and $10^4$ yrs 
(dotted). 
(b) The magnetic field strength in the pulsar wind postshock region (inside the nebula) 
as a function of time. 
(c) The differential density of synchrotron photons inside the nebula at a specific time
(the curves are marked as in figure (a)).} 
\label{fig2}
\end{figure}
Described above pulsar-supernova model is applied to a variety of high 
energy processes observed from the direction of the Cyg OB2 association. To describe 
self-consistently all the observed radiation processes, we postulate that the pulsar 
was born with the initial period $P_{\rm 0} = 2$ ms and the surface magnetic field 
$B = 6\times 10^{12}$ about $t = 2\times 10^4$ yrs ago, within or close to the dense 
core of the OB2 association. If this pulsar loses energy mainly on emission of 
electromagnetic radiation, then its present period should be $\sim 210$ ms (see Eq. 3).
A pulsar with such parameters resembles the Vela type pulsars, i.e. it should 
produce the pulsed $\gamma$-rays with luminosity comparable to the Vela 
pulsar itself or e.g. PSR 1706-44. Note that these pulsars are surrounded by the 
synchrotron pulsar wind nebulae from which the steady TeV $\gamma$-ray emission is 
observed. It is assumed that the supernova exploded close to the dense regions of the
Cyg OB2 in the medium with the density of at least $\sim 30$ cm$^{3}$ (Butt et al.~2003), 
and the initial velocity of the bulk mass of the supernova 
envelope is 2000 km s$^{-1}$ and its mass is 4 M$_{\odot}$ (as derived for the Crab 
Nebula, Davidson \& Fesen~1985). For these parameters of the pulsar and the supernova, 
the velocity of the supernova front wave after $2\times 10^4$ yrs is $\sim 100$ km 
s$^{-1}$. Such a low velocity front wave can be difficult to observe due to its 
distortions in the interactions with the clumpy high density medium of the Cyg OB2 
association. In fact, the supernova remnant shell-like structure has been revealed near the 
TeV source region in the CO, HI, and radio emission maps 
(Butt et al.~2003). The average density of matter inside this shell ($\sim 30$ cm$^{-3}$) 
is much smaller than the average density of matter in the core of the Cyg OB2 equal to 
$\sim 300$ cm$^{-3}$, calculated from the total mass of the Cyg OB2 $\sim 10^5$ M$_{\odot}$ 
(Butt et al.~2003) and its core radius $\sim 15$ pc (Kn\"odlseder~2000).

In the forthcoming sections we argue that the GeV and TeV $\gamma$-ray 
emission from the Cygnus region can be due to the Vela type pulsar and its nebula, and 
the excess of $\sim 10^{18}$ eV cosmic rays is due to the neutrons from disintegration 
of very relativistic nuclei which were accelerated soon after the pulsar formation and 
are still present inside the high density region of the Cyg OB2 association.

\section{3EG J2033+4118 - a Vela type pulsar}

The spectrum of 3EG J2033+4118 was
originally described as consistent with a simple power law with the index of $1.96\pm
0.10$, but a double power law or a power law with the exponential cut-off give better fits
(Bertsch et al.~2000, Reimer \& Bertsch~2001). This source is classified as being a 
non-variable (see discussion in Butt et al.~2003), in contrary to 
its earlier classification as a variable by Mukherjee et al.~(2003).
3EG J2033+4118 is believed to be located within the Cyg OB2 
association (Aharonian et al.~2002, Mukherjee et al.~2003, Butt et al.~2003).

Interestingly, the $\gamma$-ray emission of this source resembles the spectra
of the middle aged pulsars (the break at $\sim 1$ GeV and the sharp cut-off at a few tens 
of GeV, required in the spectrum of 3EG J2033+4118 by the level of TeV 
emission observed by the HEGRA telescope).
Therefore, we suggest that the $\gamma$-ray source 3EG J2033+4118 is caused by a yet 
undiscovered Vela type pulsar within the Cyg OB2 association. In fact, also the level of 
the $\gamma$-ray 
emission from 3EG J2033+4118 is very close to that observed from the Vela type pulsars. 
Fig.~3 shows the spectra of the Vela pulsar (V), after its re-normalization from 
the Vela pulsar distance $\sim 500$ pc to the distance to the Cyg OB2 $\sim 1.7$ kpc, 
and the PSR 1706-44 (P), which is at a similar distance as the Cyg OB2.
The spectrum of the Vela pulsar is measured by the EGRET (Thompson et al.~1997), COMPTEL
(Sch\"onfelder et al.~2000), OSSE (Strickman et al.~1996), 
and RXTE (Harding et al.~2002). The spectrum of PSR 1706-44 is measured by the EGRET
(Thompson et al.~1996) and constrained by the upper limits on the pulsed emission 
in the X-rays by the ROSAT and ASCA (Becker et al.~1995, Finley et al.~1998).
It is clear that the $\gamma$-ray spectra of these sources are very similar. The present 
parameters of the pulsar inside the Cyg OB2 can be like those ones in the example 
considered in the previous section, i.e. the age of $2\times 10^4$ yrs, the present 
pulsar period of $\sim 210$ ms, and the surface magnetic field of $6\times 10^{12}$ G. 
We encourage the searches of such order of periodicity in the $\gamma$-ray data of the
source 3EG J2033+4118.

\section{TeV J2032+4130 - a pulsar wind nebula}

As it has already been mentioned, the Vela type pulsars are surrounded by the non-thermal 
nebulae observed in the X-rays and TeV $\gamma$-rays. The present parameters of 
the pulsar, which we propose as respossible for the high energy processes in the Cyg OB2, 
are the closest to the pulsar PSR 1706-44. It lies at a similar distance, equal to 
$\sim 1.8$ kpc, has the present period of 102 ms, and the surface magnetic field of
$3.1\times 10^{12}$ G. The level of non-thermal X-ray emission from the nebula
around PSR 1706-44, observed by the ROSAT and ASCA (Becker, Brazier \& Tr\"umper~1995, 
Finley et al.~1998), is close to the level of diffuse X-ray emission observed 
by the Chandra from the direction of the $\gamma$-ray source TeV J2032+4130 in the Cyg OB2 
(Butt et al.~2003).

\begin{figure*}
  \vspace{8.truecm}
\includegraphics{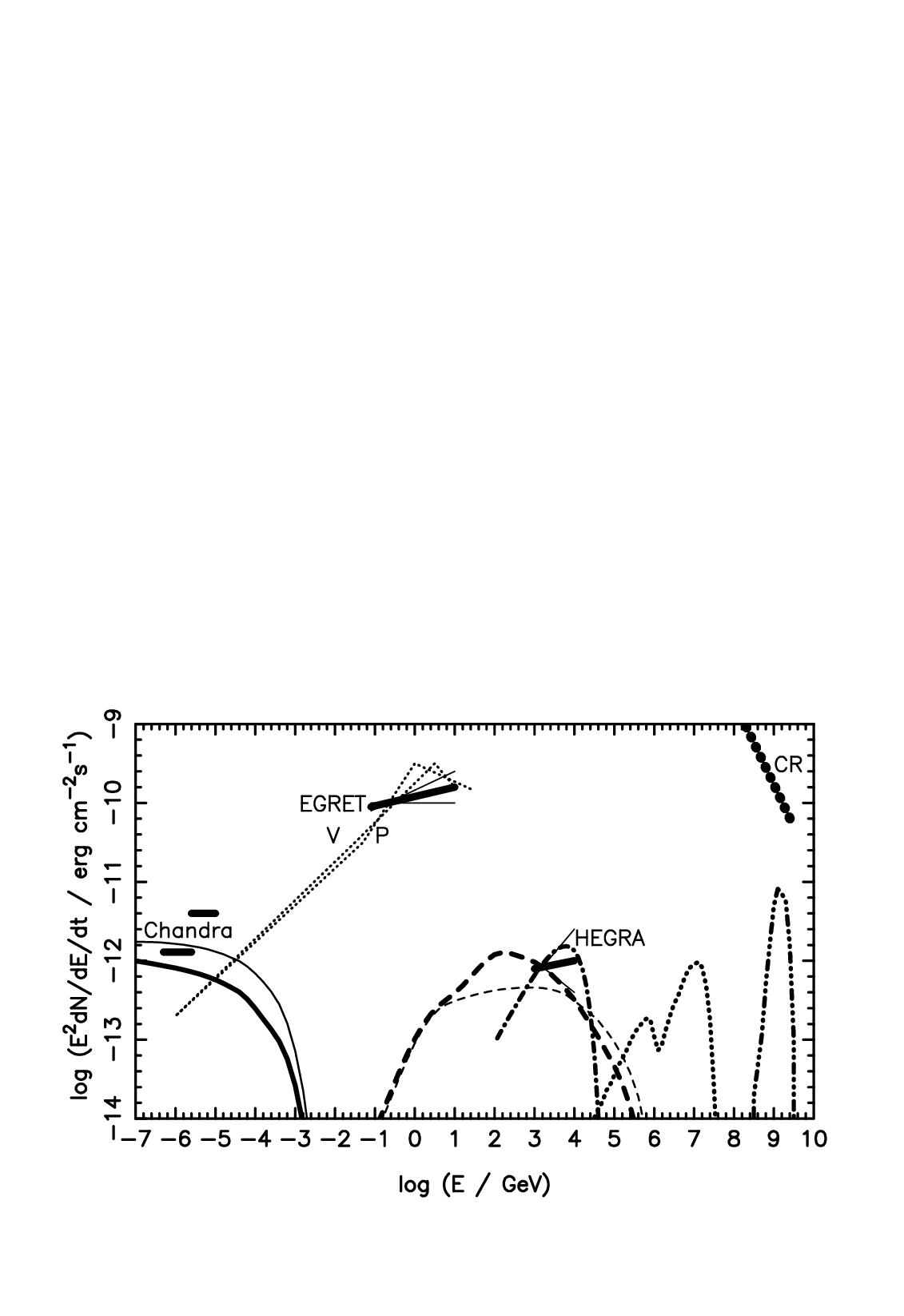}
  \caption{The broad band spectrum of the Cyg OB2 association: the diffuse X-ray emission 
(Chandra - Butt et al.~2003), the GeV $\gamma$-ray emission from the EGRET source 
3EG J2033+4118 (Hartman et al.~1999, Lamb \& Macomb~1997)), the TeV emission from the 
HEGRA source (Aharonian et al. 2002). The cosmic ray flux at EeV energies is marked by 
the filled circles (CR). The spectra of Vela type pulsars i.e., the Vela pulsar (V) and 
PSR 1706-44 (P), are shown by the thin dotted curves. The synchrotron and IC spectrum 
from the nebula around the pulsar are shown by the full and dashed curves, respectively. 
These spectra are reported for two targets of soft photons inside the nebula: (1) 
only the MBR and synchrotron photons (thin dashed curve); (2) additional infrared photons from 
the nebula (thick dashed). The spectrum of $\gamma$-rays from collisions of hadrons 
with the matter inside the nebula is shown by the dot-dashed curve. The spectra of 
neutrons (included their decay during propagation from the distance of 1.7 kpc) and 
$\gamma$-rays, produced in hadronic collisions by nuclei captured in dense regions of 
the Cyg OB2, are shown by the dot-dot-dot-dashed and thick dotted curves, respectively.} 
\label{fig3}
\end{figure*}

The X-ray to TeV $\gamma$-ray emission from the nebula around PSR 1706-44 has 
recently been interpreted in terms of the hadronic, leptonic model for the high energy 
radiation from the pulsar wind nebulae (Bednarek \& Bartosik~2003). 
The relatively low level of X-ray emission from this nebula (as compared to e.g. the 
Crab Nebula) is explained if the efficiency of positron acceleration by the nuclei 
injected by the pulsar is relatively low $\xi\approx 0.08$. On the other hand, in order 
to explain the level of the TeV $\gamma$-ray emission, an additional target of infrared  
photons inside the nebula is required, apart from the low energy synchrotron photons 
and the MBR (see Fig.~8 in Bednarek 
\& Bartosik~2003). The $\gamma$-rays, produced in collisions of hadrons with the matter of 
the nebula, do not contribute essentially to the TeV $\gamma$-ray spectrum observed from 
this nebula due to the relatively low density of matter around PSR 1706-44. 

In contrast to the PSR 1706-44 and its nebula, which were modeled by the pulsar 
with the initial period of 15 ms, here we postulate that the pulsar in the Cyg OB2 was 
born in a recent supernova explosion of one of the massive stars 
with a much shorter initial period, of the order of 2 ms. 
This supernova exploded in a much denser surrounding, estimated at 
$\sim 30$ cm$^{-3}$ by Butt et al.~(2003), than that one considered for the progenitor of  
the PSR 1706-44. Note that the average density of the Cyg OB2 association is significantly 
higher than the above mentioned value. This can be explained by the sweeping effect of the 
expanding strong wind created by the massive, pre-supernova star.
Because of the shorter initial period of the pulsar and the higher density medium, 
the nebulae around the PSR 1706-44 and the pulsar in the Cyg OB2 should differ 
in some aspects.

In terms of the hadronic-leptonic model discussed by Bednarek \& Bartosik~(2003), 
we calculate the synchrotron and the IC
spectra produced by leptons, which gain energy in the resonant interactions with 
nuclei, and the $\gamma$-ray spectra from decay of pions, produced in the interactions 
of nuclei with the matter inside the nebula. An example equilibrium spectra of leptons 
and nuclei inside the nebula $2\times 10^4$ yrs after supernova explosion are shown in 
Fig.~1. In these calculations it is assumed that the surface magnetic 
field of the pulsar born in the Cyg OB2 association is equal to $6\times 10^{12}$ G, and 
its initial period of 2 ms. Note, that the rotational energy lost by such a pulsar contributes 
significantly to the total energy budget of the expanding nebula.

The nuclei with the equilibrium spectrum interact with the matter of the PWNa
producing $\gamma$-rays via decay of pions. These $\gamma$-ray spectra  
are calculated by applying the scaling break model for hadronic interactions proposed by
Wdowczyk \& Wolfendale~(1987). The $\gamma$-ray spectra emitted $2\times 10^4$ yrs
after supernova explosion are shown in Fig.~3. Note that the level of this $\gamma$-ray 
emission is consistent with the 
flux reported by the HEGRA group. The situation with the synchrotron and IC spectra 
produced by leptons is more complicated since the low energy photon target inside the 
nebula is not precisely known. First, we consider the case when the soft photon field is 
created only by the microwave background radiation (MBR) (the energy density of 
synchrotron photons is already negligible for a $2\times 10^4$ years old 
nebula). The results of calculations of the synchrotron and IC spectra from the 
nebula with the age of $2\times 10^4$ yrs are shown in Fig.~3, assuming that the 
efficiency for lepton acceleration by the nuclei is equal to $\xi = 0.05$.
Such low efficiency is required in order not to exceed the limit on the diffuse X-ray 
emission from the Cyg OB2 observed by the Chandra telescope (Butt et al.~2003).
Next, we increase the soft photon field inside the nebula by adding the  
infrared photons produced by the gas inside or around the nebula.
The temperature of this radiation is 100 K and the average energy 
density is 5 times higher than that of the MBR. The synchrotron and IC spectra calculated 
with the presence of this additional infrared target are shown by the thick dashed curve 
in Fig.~3, using this same efficiency of lepton acceleration as before.
It is evident that only the model with the additional infrared target inside the nebula 
can produce IC $\gamma$-ray flux on the level consistent with reports by the HEGRA group.
However, this emission has a much steeper spectrum above 1 TeV, spectral index $\sim 2.5$,
than that one from the decay of pions. 
Therefore, the observations of TeV J2032+4130 at energies below 1 TeV should decide 
which processes, hadronic or leptonic, play the main role in this source. 
The comparable contribution of both processes to the $\gamma$-ray flux at $\sim 1$ TeV 
requires fine tuning of the acceleration efficiency of leptons by the nuclei to the 
value mentioned above.

\section{Cygnus CR excess - a remnant of very young pulsar}

The pulsar with the initial parameters: $P_{\rm o} = 2$ ms and $B = 6\times 10^{12}$ G,
can generate electric potentials in which in principle nuclei can be accelerated 
above $\sim 10^{18}$ eV per nucleon (see Eqs.~1 and 2). Nuclei with such energies will
soon leave the nebula around the pulsar and propagate in the surrounding medium. 
The model discussed above for the evolution of the PWNa
allows us to calculate the radius of the nebula and the magnetic field inside the 
nebula at an arbitrary time (see Fig.~2b and also Fig.~1 in Bednarek \& Bartosik~2003).
Nuclei injected inside the nebula at the time, $t_{\rm acc}$, escape from it at 
the time, $t_{\rm esc}$, if their diffusion distance, $R_{\rm dif}$, in the magnetic 
field of the nebula is equal to the dimension of the nebula, $R_{\rm Neb}$, at the 
time $t_{\rm esc}$. When estimating the escape time of nuclei from the nebula we take 
into account the structure of the magnetic field inside the nebula (its dependence on 
the radius), described by the model of Kennel \& Coroniti~(1984). Also the time evolution 
of the magnetic field at the pulsar wind shock is taken into account. 
The diffusion of nuclei through the nebula is determined by the diffusion coefficient
\begin{eqnarray}
D_{\rm dif} = (c R_L/3)^{1/2}, 
\end{eqnarray}
\noindent
where $R_L = E_{Fe}/eZB(t)$ is the Larmor radius of nuclei, $B(t)$ is the magnetic field 
inside the nebula at the time {\it t}. The distance travelled by the nuclei is then 
obtained by integration over the time
from the moment of injection up to the escape (see for details Bednarek \& 
Protheore~2002),
\begin{eqnarray}
R_{\rm dif} = \int_{t_{\rm acc}}^{t_{\rm esc}}\sqrt{{{3D}\over{2t'}}}dt'. 
\end{eqnarray}
\noindent
The above mentioned condition, $R_{\rm dif} = R_{\rm Neb}$, allows us to determine the
moment when nuclei with different energies and injection time escape from the nebula. 
The spectra of nuclei which managed to escape from the nebula during the first
$2\times 10^4$ yrs after explosion of supernova are shown in Fig.~4, assuming the  
initial parameters of the pulsar and nebula as considered above. 
Our calculations show that nuclei with the Lorentz factors above $\sim 10^{6}$ should
diffuse out of the nebula into the surrounding medium during the first $2\times 10^4$ yrs.

We consider the situation in which the pulsar (and its nebula) is formed inside 
or close to the high density regions of the Cyg OB2 association. The nuclei with the 
highest energies, Lorentz factors $\sim 10^9$, which escaped into the galactic medium 
with the magnetic field of 
$\sim 5\times 10^{-6}$ G, have the  Larmor radii of $\sim 200$ pc, and diffuse during 
$2\times 10^4$ yrs up to $R_{\rm dif}^{\rm ISM}\sim  600$ pc from the Cyg OB2. 
It is likely that 
a part of these nuclei can be captured by dense regions for $\sim 10^4$ years.
The magnetic fields in dense molecular clouds are much stronger than 
in the Galactic medium, and can reach the values of $\sim 10^{-4}-10^{-3}$ G 
(e.g. Crutcher~1999).
In such magnetic fields, the Larmor radius and the diffusion distance of nuclei with the 
Lorentz factors $\sim 10^9$ are $\sim 10$ pc and $\sim 140$ pc
during the time of $2\times 10^4$ years. These distances are close to the radius of the 
central core in the Cyg OB2 estimated on $R_{\rm OB2}\sim 15$ pc and to its diameter of 
the equal to $\sim 60$ pc. 
Therefore, since the pulsar is close to the Cyg OB2, we estimate that up to  
$\xi \sim (R_{\rm OB2}/R_{\rm dif}^{\rm ISM})^{-3}\approx 10^{-5}$ of all injected 
nuclei can be captured by the nearby Cyg OB2. 

The spectra of nuclei which escaped from the PWNa are shown in Fig.~4. Nuclei with 
different mass numbers were produced in 
fragmentation of heavy nuclei (the iron group), occurring during their collisions with 
the matter of the expanding supernova remnant during the first several years 
after explosion, when the column density of matter was still large. Note that two 
groups of heavy nuclei can be distinguished in the total spectrum 
(see Fig.~4). The higher energy heavy nuclei escape from the 
supernova envelope just after their injection into the nebula by the pulsar
since their Lorentz factors are large enough that corresponding 
Larmor radii are comparable to the dimension of the 
nebula at that time. On the other hand, the lower energy heavy nuclei escape
from the nebula at a later time ($\sim 10^3$ years), after suffering significant 
adiabatic energy losses due to the expansion of the nebula.

We calculate the flux of neutrons produced in collisions of nuclei with the matter inside
the Cyg OB2 association. The average density of matter in the Cyg OB2 is estimated at 
$\sim 300$ cm$^{-3}$, for the core radius of the Cyg OB2 equal to 15 pc and its total 
mass of $\sim 10^5$ M$_{\odot}$. 
It is assumed that in a single collision, the nuclei are split into two lower mass
stable nuclei with the rest of nucleons released free. The energies of these nucleons are
reduced by a factor of two in comparison to energies of nucleons in primary nuclei. The 
flux of neutrons produced in this process is shown in Fig.~3. 
Interestingly, the neutron flux calculated in terms of the model discussed here, is 
consistent with the excess of cosmic ray events at energies 
$\sim 10^{18}$ eV, reported from the Cygnus region by the AGASA group
equal to $\sim 8\%$ of the cosmic ray flux (Hayashida et al.~1999b). 

\begin{figure}
  \vspace{6.5truecm}
\includegraphics{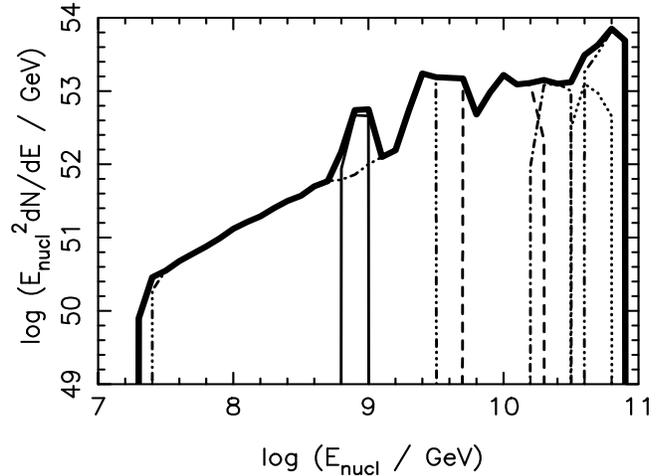}
  \caption{The spectrum of nuclei (thick curve) which escaped from the pulsar wind 
nebula into the surrounding medium. The spectra of nuclei with different mass numbers
are shown by: the thin full curve (protons), dashed curve (A=2-10), dot-dashed curve 
(A=11-20), dotted curve (A=21-40), and dot-dot-dot-dashed curve (A=41-55).
It is assumed that the pulsar was born $2\times 10^4$ yrs ago with the 
initial period of 2 ms and the surface magnetic field of $6\times 10^{12}$ G, and the
initial velocity of the supernova remnant is 2000 km s$^{-1}$ and its mass is 
4 M$_{\odot}$.} 
\label{fig4}
\end{figure}
\section{Conclusion}

We have shown that different types of high energy phenomena observed from the direction 
of the Cyg OB2 association can be self-consistently explained by the model of energetic 
pulsar created in the recent explosion of one of the massive stars.
The pulsar should have the initial period close to 2 ms. For the present age of 
$2\times 10^4$ yrs, its period should be close to $\sim 210$ ms, if its surface magnetic field is 
$6\times 10^{12}$ G. Such a pulsar belongs to the family of the so-called
Vela type pulsars, producing pulsed $\gamma$-rays with the intensity comparable to the 
EGRET source, 3EG J2033+4118, observed in the direction of the Cyg OB2. In fact, the 
spectral features of 3EG J2033+4118 (possible
break in the spectrum, required high energy cut-off) are characteristic for the spectra 
of the Vela pulsar and the PSR 1706-44. Therefore, we encourage the searches of the 
hundred millisecond periodicity in the $\gamma$-ray sources reported in the direction of 
the Cyg OB2.

The pulsar in the Cyg OB2 should also be surrounded by the pulsar wind nebula, as 
observed in the case of other Vela type pulsars. We propose that the TeV $\gamma$-ray 
source reported by the HEGRA group is due to the $\gamma$-rays produced by leptons 
and/or hadrons accelerated by the PWNa inside or close to the Cyg OB2. Therefore, in spite 
of the close physical relation, the TeV $\gamma$-ray emission from the Cyg OB2 (produced in 
the PWNa) do not need to connect smoothly with the GeV $\gamma$-ray emission 
(produced in the pulsar magnetosphere).

We also suggest that the cosmic ray excess, observed from the Cygnus region by 
the AGASA experiment, is also physically related to this energetic pulsar. The highest 
energy nuclei, injected by the pulsar soon after its formation, escape from the PWNa.
A part of these nuclei can be captured by the strong magnetic fields in the high density
regions of the Cyg OB2. These nuclei collide with the matter of the Cyg OB2, injecting neutrons 
which create the observed excess of cosmic rays in the direction of the Cyg OB2.

The $\gamma$-rays from the Cyg OB2, produced in collisions of hadrons with the
matter, should also be accompanied by the comparable fluxes of neutrinos. However, 
the predicted muon neutrino event rate in a 1 km$^2$ detector of the IceCube type during 
a 1 yr observation is rather low 
(based on the calculations of the likelihood of detecting neutrinos by such a detector
obtained by Gaisser \& Grillo~(1987), and the absorption coefficients of neutrinos in the 
Earth derived by Gandhi~(2000)). We estimate that hadrons captured in the dense regions of 
the Cyg OB2 should give between $\sim 0.14$ to $\sim 0.05$ neutrino events per year, 
for the 
source observed close to the horizon (not absorbed by the Earth) and from the Nadir 
direction (absorbed by the Earth), respectively.
The hadrons confined in the PWNa should produce $\sim 0.48$ ($\sim 0.46$) events per yr 
for the source located at the horizon (the Nadir). The spectra of neutrinos produced by 
hadrons in the PWNa and the Cyg OB2 are shown in Fig.~5. Neutrino spectra  
produced in the Cyg OB2 are above the atmospheric neutrino background expected within 
1$^{\rm o}$ of the source but they are below the sensitivity of the 
1 km$^2$ neutrino detector. On the other hand, the neutrino spectra produced inside 
the PWNa are closer to the sensitivity limit of the 1 km$^2$ detector but they are 
significantly below the atmospheric neutrino background. Therefore, we do not
predict any detection of neutrinos from the direction of the Cyg OB2 association.

\begin{figure}
  \vspace{6.5truecm}
\includegraphics{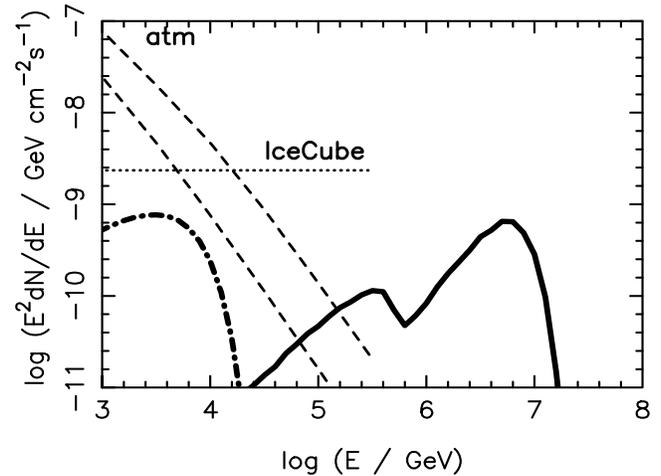}
  \caption{The spectrum of muon neutrinos, produced by hadrons captured in the 
high density regions of the Cyg Ob2 association (full curve) and produced
by hadrons inside the PWNa (dot-dashed curve), are compared with the atmospheric 
neutrino background (dashed curves) within $1^{\rm o}$ of the source for the zenith 
and azimuth directions (Lipari~1993). The 3-yr sensitivity of 
the IceCube detector is marked by the dotted line (Hill~2001).} 
\label{fig5}
\end{figure}

Finally, we conclude that the presence of at least one Vela type pulsar in the Cyg OB2 
association is quite likely on statistical grounds, since the supernova rate in this 
association should be one per a few tens of hundreds years on average
(the Cyg OB2 contains about 100 O type stars with the lifetime of a few milion years).

\section*{Acknowledgements}
I would like to thank the referee, F. A. Aharonian, for useful comments.
This work is supported by the Polish KBN grants No. 5P03D02521 
and PBZ KBN 054/P03/2001.

\end{document}